# Poisoning of magnetism in silicon doped with Re, caused by a charge transfer from interstitials to substitutionals, by means of the self-interaction corrected density-functional approach

Małgorzata Wierzbowska
*Institute of Theoretical Physics, Faculty of Physics,
University of Warsaw, ul. Hoża 69, 00-681 Warszawa, Poland*
(Dated: July 12, 2012)

The self-interaction corrected density-functional calculations are performed for Re impurities and their pairs in silicon. Rhenium ions form in the host crystal not very tight pairs, with impurities separated by one Si atom or by a distance close to two silicon bonds. Comparison of formation energies for various pairs of substitutionals, interstitials, and mixed-site impurities favours the last type. Electron transfer from the interstitial into the substitutional impurity makes the both Re sites nonmagnetic, but the p-type and the n-type co-doping revives magnetism again, the latter more efficiently.

## I. INTRODUCTION

Diluted magnetic semiconductors (DMS) attracted much experimental and theoretical attention over past decades, and still remain among hotest topics in physics and chemistry, due to their potential power to be used in spintronic devices.[1–3]

Silicon is a basic material for electronic devices, and therefore it is also a popular host for the DMS, similarly as germanium, doped typically by transition metals from the third row of periodic table.[4–7] Only recently, we got interest in studies of silicon doped with the fith-row element, namely rhenium.[8] Interestingly, amorphous silicon doped with rhenium has been experimentaly observed to be ferromagnetic at room temperature, however this phenomenon disappears when samples stay in ambient conditions for a few months.[9]

In the previous work,[8] we investigated the electronic structure of single Re impurity in various sites in the crystal, the substitutional and a few interstitial sites, by means of the density functional theory[10] (DFT) and the DFT+U scheme.[11] We already know that the single substitutional site, with magnetization 1 $\mu_B$, is energetically favoured over the interstitial site, and this fact makes the system more stable against the diffusion of impurities. The interstitial impurity is nonmagnetic in the DFT and the DFT+U description. Both methods are unable to treat properly the self-interaction effects in semiconductors, in effect, the hybridization is overestimated and the silicon gap is underestimated, and the $d$-states of unbound Re fall into the conduction band. Moreover, it has been found that silicon doped with the substitutional Re impurities is a half-metal in the minority-spin channel.

Application of self-interaction corrected approach opens new possibilities to study deeper the system of our interest. Enlarging the host gap causes a movement of the interstitial impurity states from the conduction band into the gap. This, in turn, makes the interstitial impurity magnetic, and interestingly, the whole system is a half-metal in the majority-spin channel, in contrast to the substitutional doping.

Thanks to the new approach, one can focus on neutral and charged states of both type impurities, substitutionals and interstitials, and their pairs. It is found that a charge transfer from the interstitial to the substitutional site makes both impurity centers nonmagnetic. Fortunately, the p-type and the n-type doping change the situation and the magnetism is recovered. All substitutional pairs, except the two most close pairs, are magnetic. In contrast, among interstitial pairs, there are more geometric configurations which are nonmagnetic. All charged-neutral mixed pairs (substitutional-interstitial) are nonmagnetic, and unfortunately they have lower formation energies in comparison to the pairs formed by impurities of one kind. At the end, the calculated cells have been charged with electrons or holes. In most situations, except pure interstitials, the donor co-doping would be magnetically prefered, although, it is energeticaly a bit less favoured. This is quite surprising conclusion for DMS, which usually are more magnetic under the acceptor co-doping, in contrast to the diluted magnetic oxides (DMO) which prefer the n-type co-doping.[12]

Finally, one should introduce the theoretical method used in the calculations performed here - this is the pseudopotential self-interaction corrected (pSIC) scheme proposed by Filippetti and Spaldin,[13] and developed further in a review paper.[14] Among many SIC methods formulated over decades, this is probably the newest scheme, and its simplicity drives it very popular.[15–17]

The paper is organized as follows: in the next section the computational details are given, in section III the results for neutral and charged single Re and one mixed long-distance pair are presented, in section IV the neutral pairs of all kinds are discussed, in section V the charged pairs are described, and the summary is in section VI.

## II. COMPUTATIONAL DETAILS

Calculations in this work have been done within the DFT method[10] corrected for the self-interaction in the way proposed by Filippetti et al.[13] I implemented[18] the pSIC equations into the plane-wave QUANTUM ESPRESSO code,[19] within the ultrasoft Vanderbilt pseudopotentials[20] (USPP) scheme. The generalized gradient corrected (GGA) functional of the Perdew-Wang type[21] (PW91) has been chosen for all calculations.

The Re pseudopotential, as in the previous work,[8] has been generated for the valence configuration $5d^5 6s^2$ including the nonlinear core correction,[22] and the Si pseudopotential was of the valency 4. Both pseudopotentials are scalar relativistic, however the spin-orbit interaction was not included. For the plane-wave expansion, the 30 Ry cutoff was sufficient for the USPP. The energy cutoff for the density was 300 Ry, since the additional term for the norm conservation within the USPP scheme requires it.

Most of calculations were done in the simple-cubic supercell with 64 atoms, but bigger cells with 96 atoms have been also used, as will be pointed in the results section. As for the integration within the Brillouin zone, the k-point grid was generated using the (4,4,4) division according to the Monkhorst and Pack method,[24] and the metallic broadening[25] of 0.01 Ry has been used for a better convergence.

All calculations were performed with the silicon lattice constant of 10.32 a.u., which was optimized within the GGA method. The lattice relaxations around impurities have been found with the GGA method, and used for the single-geometry calculations whithin the pSIC scheme.

The charged impurities were calculated together with the compensating uniform charged background.[23] The formation energy of impurity in the neutral/charged cell is defined as:

$$E_F = E_T - N_{Si} \cdot \mu_{Si} - N_{Re} \cdot \mu_{Re} - N_e \cdot \mu_e, \quad (1)$$

where $N_{Re}$ and $N_{Si}$ are the numbers of Re and Si atoms in the supercell, $E_T$ is the total energy of the neutral/charged system, and $\mu_{Si}$ and $\mu_{Re}$ are chemical potentials of the reservoirs of Si and Re atoms, $\mu_e$ is the Fermi energy of the host, and $N_e$ is a number of additional electrons ($N_e < 0$ for holes). For the parameter $\mu_{Si}$ the total energy of the silicon bulk per atom is taken. For relative energies, one does not need to know $\mu_{Re}$, which however is necessary for the absolute formation energies. The chemical potential $\mu_{Re}$ might be obtained from the total energy of hcp Re metal, or from the compounds of Re and Si (for instance ReSi$_2$), or from an isolated atom. These approaches lead to very spread values of the absolute formation energies. The choice of reference $\mu_{Re}$ should be made in a connection with the experiment with which one likes to compare theoretical results. Therefore in this work, I present only the relative formation energies.

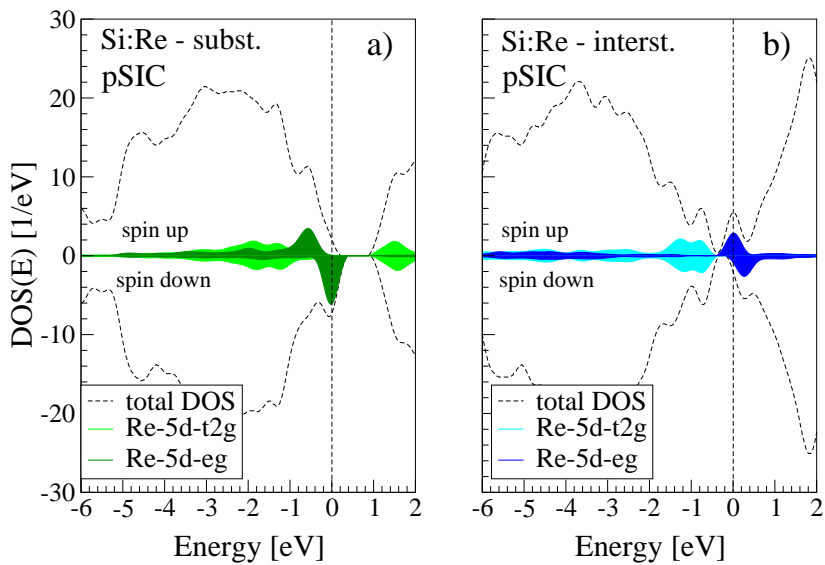

FIG. 1. (Color online) Density of states of a) the substitutional Re, and b) the interstitial Re impurities in silicon in the 64-atom neutral supercells, from the pSIC calculations. The Re-5$d$ e$_g$ and t$_{2g}$ states are scaled by two for better visualization. Energy zero is at the Fermi energy.

## III. SINGLE IMPURITIES AND THE CHARGE TRANSFER BETWEEN THE INTERSTITIAL AND THE SUBSTITUTIONAL RHENIUM IONS

In our previous work,[8] the density of states (DOS) of single impurities in the substitutional and the interstitial sites have been presented for the GGA method, and in a case of the substitutional Re also for the GGA+U method. The interstitial Re-$5d$ states fall into the conduction band within the GGA, due to underestimation of the silicon gap, and therefore this site is nonmagnetic. Here, Fig. 1 shows the DOS of Re in two sites, the substitutional (S-site) and the interstitial (I-site), obtained from the pSIC scheme. The DOS of the S-site impurity from the pSIC is similar to the GGA+U result. The interstitial-Re, however, becomes magnetic in the new method, since the Re-$5d$ states touch the conduction band from the bottom. One gets the above result due to a slight opening of the silicon gap from 0.62 eV (from the GGA) to 0.75 eV (from the pSIC), which turned out to be sufficient (the experimental gap in Si is 1.17 eV) for a proper description of Si:Re.

Interestingly, the half-metallicity is in the minority-spin channel for the S-site Re, and it is opposite, in the majority spin, for the I-site Re. Therefore, it could be plausible to characterize samples with the spin-Hall experiment.

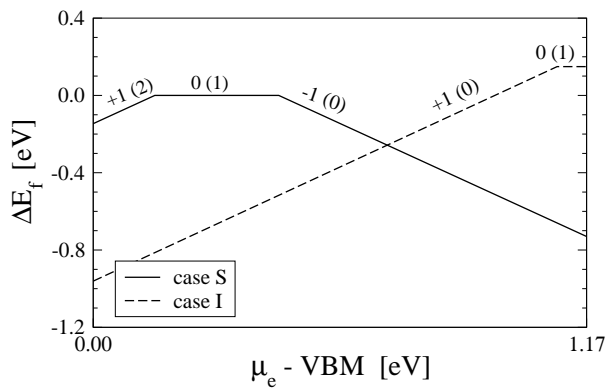

FIG. 2. Energies of the S-site and the I-site Re in silicon ($\Delta E_f$), in the neutral and charged 64-atom (65-atom for I) cells, calculated within the pSIC, and given with respect to the energy of the S-site Re. Numbers above the curves denote the charges of the supercells. Numbers in parenthesis denote the total magnetic moments.

As for the geometry relaxation, the Wyckoff positions have been optimized with the GGA and used in the pSIC calculations. The relaxation is negligible for the S-site Re (about 0.03 %) and it acts towards shortening of the Si-Re bond. In contrast, the geometry relaxation for the I-site Re is visible (about 4 %) and oriented towards prolongation of the Si-Re bond. Further relaxations of the charged cells with respect to the neutral case are negligible,[8] therefore, I do not take care of them in this work.

In the first paper about Si:Re,[8] we showed the energetics obtained with the GGA method for the S-site and the I-site Re in the neutral and the charged cells. In this work, Fig. 2 presents the corresponding states from the pSIC calculations. The formation energy of the I-site Re is still higher than that of the S-site Re, of 149 meV in the neutral cell. Additionally, it is found that the p-type co-doping for the S-site Re might be plaussible. The n-type co-doping of the I-site Re is still prohibited by a too small gap of the host.

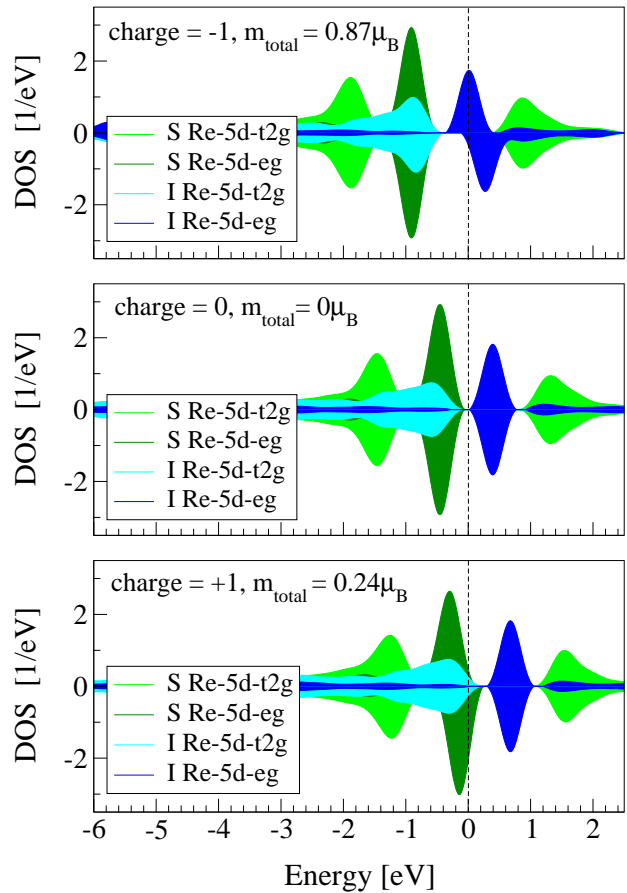

FIG. 3. (Color online) Density of states of the S-site and the I-site Re in silicon, calculated with the pSIC method, in a) the charge negative, b) the neutral, c) the charge positive 97-atom cell. Energy zero is at the Fermi energy of given supercell (Fermi energy is not the same for all supercells).

It is very interesting to note that similarly to the GGA result, also in the pSIC, the lines of the n-type doping of the S-site and the p-type doping of the I-site cross each other below the energies of the neutral cases. This observation is very important for magnetism in Si:Re, since it is energetically favourable situation for an electron transfer from the I-site to the S-site impurity in the charge neutral cells. Such charge transfer between Re-impurities kills magnetism, because after hopping of an electron, the S-site Re has completely occupied $e_g$ states and the I-site Re has completely empty $e_g$ states.

In order to better investigate this process, I consider the 97-atom cell, constructed by a prolongation of the 64-atom cell in one direction, and put two impurities, S-case and I-case, inside. The impurities are placed in the most distant separation which is possible in this cell. The DOS is drawn in Fig. 3 for a) the charge negative, b) the neutral, and c) the charge positive cells. Indeed, as expected from Fig. 2, in the charge neutral cell, both impurities are nonmagnetic with the predicted occupations of the $e_g$-states. This is a bad news for us, since the statistical content of impurities in experimental samples probably is similar for both types, S and I, due to a small difference in the formation energies. Hope comes with the p-type or the n-type co-doping, then magnetization is restored on the S-site or the I-site Re correspondingly. Nevertheless, the localization of additional holes or electrons on impurities is never 100 % in extended systems. In the case of hole-doped cells with impurities of mixed type, S and I, the total magnetization amounts to about quarter of the concentration of co-dopants or injected carriers. In contrast, in the case of electron-doped mixed pairs, the total magnetization access about 0.87 % of additional charge. Thus, the n-type doping is more efficient for magnetization. The energetic issues of all charged pairs will be given at the end of section V.

## IV. NEUTRAL PAIRS OF RE-IMPURITIES

In the following three subsections, the formation energies and magnetizations of each pair in the neutral cell will be given with respect to an energy of the most stable pair within the given group of pairs: substitutional-substitutional (SS), interstitial-interstitial (II) or substitutional-interstitial (SI). I chose the reference energy to avoid necessity of using the chemical potential of Re. The choice of $\mu_{Re}$ might be controversial since it should be close to the experiment in which Si:Re was obtained. At the end of this section, in the subsection D, a comparison of the formation energies per impurity for the neutral pairs of the SS-, II- and SI-type to the formation energies of single impurities will be given.

Since the geometry relaxation is important for the interstitial impurity and for the close SS pairs, in this work I perform the GGA relaxation prior to all pair calculations with the pSIC, except for far SS pairs. Unrelaxed pairs are denoted by "$klm$-$i$", and in such a pair, the Re-Re separation in $(x, y, z)$-direction is equal to $(k, l, m) \times a/4$, where $a$ is the lattice constant, and "$i$" means the "ideal" Si-crystal symmetry; the relaxed pairs are denoted by "$klm$".

### A. Substitutional pairs

In Table I, there are collected the pair formation energies with respect to the energy of the most stable SS pair, which was found to be the relaxed 220 pair. The relaxed 111 pair is much less stable within the pSIC. This result shows that the coupling between two Re impurities is mediated strongly by the Si-Re bond. The pairs $11\bar{3}$ and 331-i also have low formation energies, while the distant pairs, such as 444-i, will be probably rare in the experimental samples.

| pair-SS | $E_f$ | $E^{AF-FM}$ | $m_{tot}^{FM}/m_{abs}^{FM}$ | $m_{abs}^{AF}$ |
|---|---|---|---|---|
| 111 | 1.938 | -56 | 1.62/1.79 | 0 |
| 220 | 0 | -27 | 1.73/2.37 | 0 |
| $11\bar{3}$ | 0.221 | 32 | 1.96/2.23 | 2.30 |
| 331-i | 0.164 | 125 | 1.86/2.15 | 2.43 |
| 400-i | 1.138 | 39 | 1.97/2.29 | 2.00 |
| 422-i | 1.191 | 20 | 1.91/2.19 | 1.88 |
| $33\bar{3}$-i | 2.716 | 2 | 1.83/2.09 | 1.93 |
| 440-i | 0.283 | -2 | 2.00/2.43 | 2.44 |
| 444-i | 2.722 | -4 | 1.76/1.99 | 1.94 |

TABLE I. Formation energy ($E_f$, in eV) for all SS-pairs in the FM state, in the 64-atom cell, with respect to the 220 pair. The magnetization energy ($E^{AF-FM}$, in meV), and the total magnetization of the cell for the FM state ($m_{tot}^{FM}$), as well as the absolute magnetizations for the FM and the AF states ($m_{abs}^{FM/AF}$) in $\mu_B$ are also given.

Magnetizations of the tight-pair 111 and the close-pair 220 calculated in the antiferromagnetic (AF) state are reduced to zero (I obtained the paramagnetic state (PM)), while the ferromagnetic state (FM) has only slightly reduced on-site spin. In the 111 and 220 pairs, the chemical bonds Re-Re and Re-Si-Re form correspondingly. These bonds are present in the diamagnetic materials such as hcp rhenium and ReSi$_2$. The PM state in these two pairs is energetically predominant over the FM state. This, however, might be a drawback of the SIC and the DFT+U methods, which treat badly the metallic states at the Fermi level, since they support more integer occupations of the atomic orbitals. In such cases, the crystal degeneracies of the impurity band are partially lifted, due to distorted crystal symmetry by a second atom from the pair of close TM-ions, and the quasi-dip forms exactly at the Fermi level, which supports the AF-order or the PM-phase (the AF-superexchange mechanism is favoured). Most of pairs of substitutionals prefer the FM spin-alignment.

As for the optimization of cells with the SS-pairs, I obtained the largest relaxation, within the GGA, for the 220 pair (23.6 %), and weaker for the 111 (6.4 %) and the $11\bar{3}$ (3 %) pairs. For the 331 pair, the optimized separation was almost unchanged (0.2 %). All relaxations were oriented towards shortening of the Re-Re distance.

### B. Interstitial pairs

In this subsection, I optimized all cells with interstitials, running the GGA method prior to the single-geometry calculations with the pSIC scheme. In contrast to the substitutionals, the interstitials repel each other at short distances. The separations increased for the pairs: 111 (22.2 %) and 220 (0.8 %), while for a more distant pair, such as $11\bar{3}$, the Re-Re separation is slightly shorter (1.1 %). It has been checked that, the magnetic state (FM or AF) has no influence on the relaxation of the geometry of the SS-pairs. I believe that it holds also for the geometries of the II-pairs, although we should be aware of the fact that the pSIC relaxations will be probably slightly weaker from these obtained with the GGA method.

| pair-II | $E_f$ | $E^{AF-FM}$ | $m_{tot}^{FM}/m_{abs}^{FM}$ | $m_{abs}^{AF}$ |
|---|---|---|---|---|
| 111 | 3.154 | 0 | 0/0 | 0 |
| 220 | 0.617 | -31 | 1.24/1.39 | 0 |
| 11$\bar{3}$ | 0 | 0 | 0/0 | 0 |
| 331 | 0.991 | -8 | 1.69/1.93 | 1.41 |
| 400 | 2.166 | 0 | 0/0 | 0 |
| 422 | 2.074 | -19 | 1.30/1.47 | 0.58 |
| 33$\bar{3}$ | 3.524 | 0 | 0/0 | 0 |
| 440 | 0.886 | -19 | 0.96/1.12 | 0 |
| 444 | 3.849 | 3 | 1.78/2.03 | 1.66 |
| $E_f$(II-11$\bar{3}$) - $E_f$(SS-220) = -0.832 eV | | | | |
| $E_f$(II-220) - $E_f$(SS-220) = -0.215 eV | | | | |

TABLE II. Formation energy ($E_f$, in eV) for II-pairs in the FM state, in the 66-atom cell, with respect to the 11$\bar{3}$ pair. The magnetization energy ($E^{AF-FM}$, in meV), and the total magnetization in the FM state ($m_{tot}^{FM}$), as well as the absolute magnetizations for the FM and the AF states ($m_{abs}^{FM/AF}$) in $\mu_B$ are also given.

Table II collects the same information for the interstitials which has been discussed earlier for the substitutionals. The most stable II-pair is 11$\bar{3}$, while in the SS-group it was the 220 pair. Comparison of energies of the II-11$\bar{3}$ pair and the SS-220 pair shows that the interstitials are favoured by 832 meV. The II-220 pair is also prefered over 220 pair of substitutionals. This result is in contrast to stability of the single impurities, where the S-site was lower in the energy than the I-site.

One can also note, that magnetism disappears in the 111 and 11$\bar{3}$ pairs, because the Re atoms are quite close and "see" each other directly. On the other hand, magnetism remains in the II-220 pair, due to "screening" of Re ions by the Si atom placed exactly on-line with the impurities. In contrast, in the SS-220 pair, the bridging Si atom forms a triangle with the Re impurities, and the AF start leads to the nonmagnetic result for this pair.

### C. Mixed substitutional-interstitial pairs

The geometry of all mixed pairs has been optimized with the GGA. The relaxed pair separations differ from the ideal Si-crystal separations by 4.7 % for the 002 pair and -1.2 % for the 222 pair, and the distances between impurities of all other SI-pairs are almost unchanged.

Table III presents only the formation energies, within the pSIC, because all mixed pairs are nonmagnetic due to a charge transfer from the I-site to the S-site, discussed

| pair-SI | $E_f$ |
|---|---|
| 002 | 0.013 |
| 11$\bar{1}$ | 3.106 |
| 113 | 0.844 |
| 222 | 2.933 |
| 240 | 0.773 |
| 33$\bar{1}$ | 0.750 |
| 333 | 0 |
| 442 | 0.739 |
| $E_f$(SI-333) - $E_f$(SS-220) = -1.210 eV | |

TABLE III. Formation energy ($E_f$, in eV), with respect to the 333 pair, for all mixed-pairs of the substitutionals and the interstitials in the 65-atom cell.

in the first part of the results section.

The most stable mixed pair is 333, and its formation energy is 1.210 eV lower than the energy of the SS-220 pair. The formation energy of SI-002 pair is very similar to the 333 pair.

Larger stability of mixed pairs is not surprising, because of the energy gain due to the charge (and spin) redistribution.

### D. Comparison of pairs to singles

At the end of Sec. II, it has been mentioned about the ambiguity in the exact determination of the chemical potential of rhenium. This problem can be omitted in a comparison of singles to doubles, by taking the formation energies of pairs per one impurity. Presentation of such energies is enclosed in Fig. 4.

Interestingly, the mixed neutral pairs are the most stable, being lower energetically than the interstitial pairs. The double substitutionals are less privileged, in contrast to the single substitutionals. This result, together with the charge transfer from the I-site to the S-site, supports a scenario of non-magnetic experimental Si:Re samples when they are not co-doped.

### V. CHARGED PAIRS OF RE-IMPURITIES

Since the S-site impurities are electronic acceptors and the I-site Re ions are electronic donors, it is desirable to know the energetics and magnetization of the charged cells with the SS and the II pairs. Especially interesting are the mixed SI pairs in the charged cells.

Table IV presents the negatively charged SS pairs. The total magnetic moment of almost all pairs is close to 1. Only the negatively charged 444 pair is nonmagnetic, which might be due to close screening of impurity (000) by the Si atom at the 111 site which is exactly on the line connecting the Re ions. The formation energies in charged cells relative to the corresponding energies in the neutral cells, with the same geometric configurations, are very similar for all pairs except the 220 and the 440 pair. The electron doping of the 220 pair (bridged by the Si atom in a triangular geometry) is energetically much less favourable than that for the 440 pair, in which the Re ions are perfectly screened by the Si atom at the 220-site.

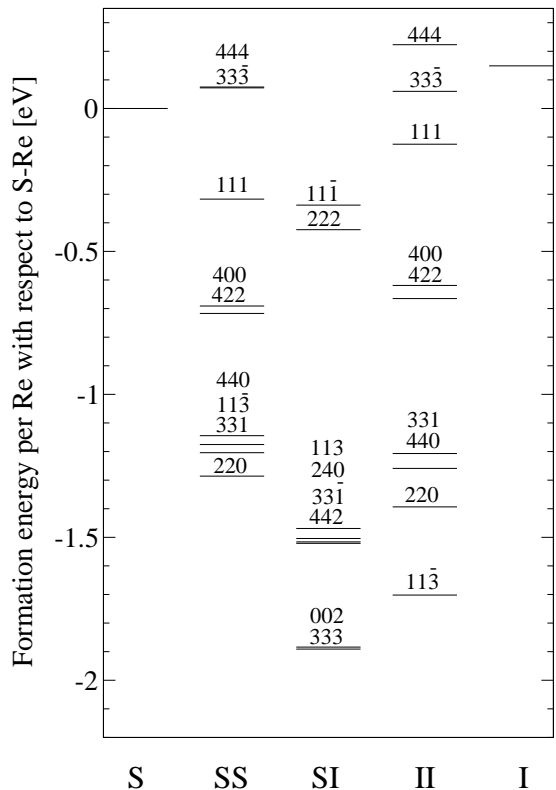

FIG. 4. Formation energies of chosen pairs SS, II and SI per one Re, with respect to the energies of single impurities S and I. Numbers above the energy levels denote the pair-orientations in the Si crystal.

| pair-SS$^{-1}$ | $E_f$ | $M_{tot}$ |
|---|---|---|
| 220 | 0.692 | 0.93 |
| 331 | 0.342 | 0.97 |
| 11$\bar{3}$ | 0.336 | 0.97 |
| 440 | 0.189 | 0.97 |
| 422 | 0.374 | 0.84 |
| 444 | 0.399 | 0 |

TABLE IV. Formation energies of the negatively charged SS-pairs with respect to the neutral SS-pairs ($E_f$, in eV), and the total magnetic moment in the 64-atom cell.

The relative formation energies for positively charged II pairs are collected in Table V. These pairs are about 1.3-1.4 eV more stable than the neutral II-pairs in the same geometric configuration. The magnetization of the pair 331 is small, about 0.3 $\mu_B$, and all other pairs are nonmagnetic. This result is not predicted from Fig. 2, since we would expect one of the interstitial ions to be slightly magnetized, in the situation when only one electron from the II-pair is removed. This fact is due to a small degree of localization of the Re-5$d$ states at the I-site ion, and partial overlap of the impurity states with the conduction band in the case of two close- or medium-distance interstitial ions.

| pair-II$^{+1}$ | $E_f$ | $M_{tot}$ |
|---|---|---|
| 11$\bar{3}$ | -1.219 | 0 |
| 220 | -1.412 | 0 |
| 440 | -1.285 | 0 |
| 331 | -1.357 | 0.26 |
| 422 | -1.382 | 0 |
| 444 | -1.435 | 0 |

TABLE V. Formation energies of the positively charged II-pairs with respect to the neutral II-pairs ($E_f$, in eV), and the total magnetic moment in the 66-atom cell.

| | charge +1 | | charge -1 | |
|---|---|---|---|---|
| pair-SI | $E_f$ | $M_{tot}$ | $E_f$ | $M_{tot}$ |
| 333 | -0.804 | 0 | 1.199 | 0.73 |
| 002 | -0.805 | 0 | 1.202 | 0.72 |
| 442 | -0.728 | 0 | 1.118 | 0.86 |
| 33$\bar{1}$ | -0.754 | 0 | 1.120 | 0.86 |
| 240 | -0.739 | 0 | 1.135 | 0.89 |
| 113 | -0.668 | 0.99 | 1.100 | 0.81 |

TABLE VI. Formation energies of the positively and the negatively charged SI pairs with respect to the neutral SI-pairs ($E_f$, in eV), and the total magnetic moment in the 65-atom cell.

Finally, the SI pairs are doped with the electrons and holes. It was expected from Fig. 3, to be plaussible to retain the magnetic moment lost in the charge transfer in these pairs. In section III, in the 97-atoms cell, the p-type doping was less efficient than the n-type doping. Table VI shows the relative formation energies for the chosen SI pairs in the charged cells and their total magnetic moments. The p-type doping is energetically prefered of about 650-800 meV, but in most pairs, the one additional hole per a pair does not localize at the Re-5$d$ states. On the other hand, the electron doping, which is less attractive energeticaly, restores magnetism in the cell for all geometric configurations.

## VI. SUMMARY

I studied silicon doped with rhenium, by means of the pseudopotential self-interaction corrected scheme. This approach corrects the silicon gap and places the impurity states in the proper position. Both the substitutional and the interstitial sites occupied by rhenium are magnetic with the localized moment of 1 $\mu_B$, and both doping sites make the crystal half-metallic, with the difference that for the substitutionals it is in the minority-spin and for the interstitials in the majority-spin channel. Interestingly, the mixed pairs of impurities, substitutionals and interstitials, make the whole system nonmagnetic due to the charge transfer from the less bound interstitial site to strongly bound substitutional rhenium. This is a bad news for the DMS, but fortunately the charged co-doping restores the magnetic properties, and in this system it is more efficient for the electron donors.

From the performed comprehensive calculations for various pairs of impurities in the neutral cells, it has been found that the lowest formation energies are obtained for the mixed substitutional-interstitial pairs and purely interstitial pairs. The pairs of Re separated by a single silicon-bond are energetically unfavourable, and the pairs close but not very tight, where impurities are connected by a single Si atom or their separation is in the similar distance, prevail. The very sparse and uniform distribution of impurities in the host is highly disliked by the system, similarly to the DMSs investigated by other authors.[3,26–29]

Almost all substitutional pairs, except the two most tight, are magnetic. In contrast, the interstitial pairs form more nonmagnetic configurations, and the mixed SI pairs are all magnetically poisoned by the aforementioned charge transfer from the I-site to the S-site.

The charged co-doping of various pairs is magnetically efficient for the n-type case, which is in contrast to most of the IV-group and the III-V group DMS, and makes the system of investigations more similar to co-doped DMOs (with TM-oxides as hosts).[12]

## VII.  ACKNOWLEDGMENTS

I thank Prof. R. R. Gałązka for showing me the experimental results and to Andrzej Fleszar for cross-checking my implementation of the pSIC method to the QUANTUM ESPRESSO code with his mixed-basis code. The calculations have been done at the Leibniz Supercomputing Centre in Munich. The work was supported by the European Founds for Regional Development within the SICMAT Project (Contract No. UDA-POIG.01.03.01-14-155/09).